# Planet RB: A Personal Contribution to a Proteomic Map of Human Retinoblastoma Protein

Razvan Tudor Radulescu

*Molecular Concepts Research (MCR), Munich, Germany*

E-mail: ratura@gmx.net

## ABSTRACT

As I compress on the canvas of a few pages here major results of my research on the retinoblastoma tumor suppressor protein (RB) spreading over the past 15 years, an exciting picture emerges on this unique host molecule which surpasses in its complexity even that of the most capable viral proteins known to date. Accordingly, RB has the potential to bind not only growth-promoting proteins such as insulin, but also to attach itself to calcium and oxygen, as well as to be secreted into the extracellular environment. Moreover, RB may exert proteolytic, antimicrobial and anti-aging activities. These condensed structure-based insights on RB are the substance of a scientific revolution I have initiated a long time ago, yet likely to gain even further speed in the years to come, thus expanding both our understanding of life at the molecular level and the possibilities for pharmacological modulation of fundamental biological phenomena, particularly in oncology and gerontology.

## A brief perspective on planet RB

I summarize here the amino acid motifs I have discovered in the crucial tumor suppressor retinoblastoma protein (RB) since 1992, suggesting that RB "mirrors" the biochemical world in that it is considerably more versatile than previously suspected, thus resembling a planet with numerous interesting peaks and valleys serving as docking sites for other molecules physically interacting with it or for its self-association. The following table gives an overview of my findings.

| Human RB residue number | Putative or proven role[Ref.] |
|---|---|
| 1-26 | signal peptide[1] |
| 3-5 | antineoplastic & anti-aging through block of insulin binding to its receptor[2] |
| 13-18 | amyloidogenic hexa-alanine counteracting (prion-associated) neurodegeneration[3,4] |
| 30-36 | calcium binding site[5] |
| 142-145 | IgG1 constant region-like fragment[6] |
| 148-152 | oxygen-mimicking peptide[7] |
| 254-257 | ß-endorphin COOH-terminus-like[8] |
| 276-280 | LXCXE-like anti-apoptotic cell survival amino acid motif[9] |
| 649-654 | insulin-binding and thus antineoplastic[10-23] as well as anti-aging[24-25]; furthermore, antimicrobial[6] & anti-inflammatory[6] |
| 673-678 | insulin-degrading enzyme (IDE) zinc binding site-like sequence[26] |
| 675-678 | oxygen binding site[27] |
| His673, Asp718 & Ser811 | serine protease catalytic triad-like[28] |

Last but not least, I have unveiled that RB contains 10 potential glycosylation sites (29) consistent with its previously proposed roles in the extracellular environment (30). This plethora of anticipated functions underscores that RB is an unique protein that remains to be further explored along these lines and beyond.

## References

1. **Radulescu, R.T.** 2004. Signal peptide-like sequence in retinoblastoma protein (RB): the signature for the secretion of a nuclear tumor suppressor. Logical Biol. **4:**81-83.
2. **Radulescu, R.T.** 2003. Potential of retinoblastoma protein to block insulin receptor activation by insulin: structural and experimental clues to a novel anti-dogma on a dual inhibition of cancer and ageing. Logical Biol. **3:**40-42.